\pdfoutput=1
\documentclass[aps,prl,twocolumn,superscriptaddress]{revtex4-2}

\usepackage{graphicx}
\usepackage{color,braket,bm,amsmath,multirow}

\begin{document}

\title{Giant Rydberg Excitons Delineated as Relativistic Bound States}
\author{Mitsuyoshi Takahata}\email{takahata@aoni.waseda.jp}
\affiliation{Department of Physics, Kyoto University, Kitashirakawa-Oiwake-cho, Sakyo-ku, Kyoto 606-8502, Japan} 
\author{Ikuko Akimoto}
\affiliation{Department of Material Science and Chemistry, Wakayama University, Sakaedani 930, Wakayama 640-8510, Japan}
\author{Nobuko Naka}\email{naka@scphys.kyoto-u.ac.jp}
\affiliation{Department of Physics, Kyoto University, Kitashirakawa-Oiwake-cho, Sakyo-ku, Kyoto 606-8502, Japan} 

\date{\today}

\begin{abstract}
We examined the energy-momentum relationship of the valence band of Cu$_2$O, 
which has gained interest as an emerging solid host for giant Rydberg states. 
Our findings reveal a nonparabolic dispersion, 
described using the rest mass (0.570 times the free electron mass) and rescaled light speed ($1.6\times 10^5$ m/s), 
which draws parallels with relativistic theory for high-energy particles. 
We propose a straightforward expression for the bound-state energies of Rydberg excitons with finite quantum defects, 
whose energy scheme deviates from the hydrogenic series.
This formulation identifies relativistic kinematics as the physical origin of quantum defects, 
offering a streamlined basis for future studies on complex many-body correlations.
\end{abstract}

\maketitle

\begin{figure}[b]
\centering
\includegraphics[width=8.5cm, clip]{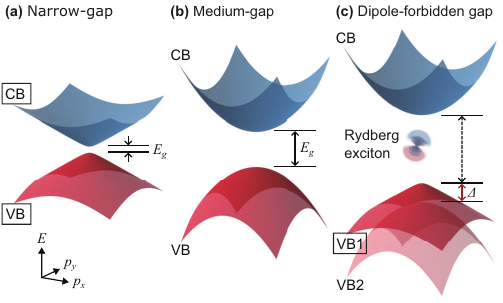}
\caption{
Schematic of $E(p)$ in (a) narrow, (b) medium, and (c) dipole-forbidden gap materials. 
Nonparabolic bands with relativistic dispersion are highlighted by box lettering. 
$E_{\rm g} $ and $\Delta$ denote the energy gap and spin-orbit energy, respectively.
}
\label{Figure1}
\end{figure}

In Dirac's relativistic theory \cite{Landau}, the relationship between the kinetic energy $E$ and momentum $p$ of high-energy particles is expressed as $E(p)=$$\sqrt{m^2c^4+p^2c^2}-mc^2$ using the particle's rest mass $m$ and the speed of light $c$. 
In solid-state systems, a mathematical resemblance to the Dirac equation is known for the two-band Hamiltonian of narrow-gap semiconductors (NGS) \cite{Kane}. 
The energy dispersion for electrons in NGS [Fig. 1(a)] is described as \cite{Zawadzki} 
\begin{equation}
E(p)=\sqrt{\left(E_{\rm g}/2\right)^2+p^2E_{\rm g}/(2\tilde{m})}-E_{\rm g}/2,
\label{semicon}
\end{equation}
where  $\tilde{m}$ is the effective mass and $E_{\rm g}$ is the energy gap between the conduction band (CB) and valence band (VB). 
Indeed, the two dispersion relations are equivalent under the correspondence of $mc^2\Leftrightarrow E_{\rm g}/2$ and $m\Leftrightarrow \tilde{m}$ \cite{Zawadzki}.
When $\tilde{m}$ approaches zero owing to strong renormalization by the periodic lattice potential in NGS, the electrons exhibit a linear dispersion akin to high-velocity particles in the relativistic regime. 
This prominent effect gives rise to dramatic phenomena, such as the anomalous Hall effect in topological insulators and ultrahigh mobility of massless fermions in Dirac and Weyl semimetals \cite{topo}. 

Conversely, semiconductors with a wider energy gap represent non-relativistic electrons with parabolic dispersion $E(p)=p^2/(2\tilde{m})$ in the effective mass approach [Fig. 1(b)].
As an exception, {\it nonparabolic} dispersion is known in the oxide semiconductor Cu$_2$O with $E_{\rm g}=2.17$ eV [Fig. 1(c)].
In crystals with inversion symmetry (such as Cu$_2$O), the top valence band (VB1) is not coupled to the bottom conduction band (CB) of the same parity through electric-dipole interactions.
Instead, VB1 is coupled to the lower-lying valence bands (VB2, shown with omitted degeneracy), separated by a spin-orbit splitting of $|\Delta|=0.13$ eV.
As a result of this coupling, VB1 deviates from a parabola and exhibits a quasilinear, that is, nonparabolic dispersion of the relativistic type (see Supplemental Material I \cite{SM,Thesis}). 

When electrons are excited from the valence band to the conduction band, the Coulomb force pairs the electrons in the conduction band with positively charged holes from the valence band.
A giant Rydberg exciton \cite{Kazimierczuk2014} is a notable example of such quasiparticle bound states, excited to a large principal quantum number $n$, that partly retain the valence band characteristics.
The energy level spectrum is expressed as \cite{Kazimierczuk2014,Seaton1966,Julian,Kim} 
\begin{equation}
E_{n,\ell}=-R/(n-\delta_{n,\ell})^2
\label{Ryd}
\end{equation}
to a good approximation, where $R$ denotes the Rydberg energy and $\ell$ is the orbital angular momentum.
Here, the quantity $\delta_{n,\ell}$, called the quantum defect, provides a phenomenological explanation of the binding energies. 
In multi-electron atoms, this deviation from Bohr's hydrogen model is attributed to the various degrees of valence electron screening by the inner-shell electrons. 
For Rydberg excitons, theory \cite{Schone2016,Coupled,Scaling,Ertl2026,Krueger} 
predicts that $\delta_{n,\ell}$ originates from the nonparabolicity of the valence bands. 
However, there have been no quantitative experimental data with sufficient accuracy regarding the band dispersion of Cu$_2$O, and the Rydberg series has not yet been discussed in relation to relativistic theory.

This Letter reports a quantitative evaluation of the nonparabolicity of VB1 in Cu$_2$O and its impact on giant Rydberg excitons.
We directly observed the effective mass variation through the time-resolved cyclotron resonance of holes under our original photoexcitation at various lattice temperatures to determine the dispersion near the center of the Brillouin zone (BZ), 
as opposed to the entire BZ probed by angle-resolved photoemission spectroscopy \cite{ARPES}.
We describe Rydberg excitons using the relativistic dispersion relation, which enables the extraction of quantum defects 
based on the Dirac equation for a relativistic particle. 

While the hydrogen-like model has traditionally been used for shallow impurity levels in materials such as silicon \cite{Silicon}, 
deviations from the hydrogenic series with finite quantum defects in a solid system have recently been identified for Rydberg excitons in Cu$_2$O \cite{Kazimierczuk2014,Scaling,PLE,Rogers}.
Analogous to Rydberg atoms \cite{atom1,atom2,atom3}, these highly excited quasiparticles have properties distinct from those of the ground state.
Their strong dipole-dipole interactions, which lead to the Rydberg blockade \cite{Kazimierczuk2014,asymmetry}, 
are relevant to electromagnetic-field sensing \cite{sensing} and quantum information applications \cite{quantuminformation,Bernien2017}. 
Their giant Kerr nonlinearities \cite{selfKerr, microwave}, confinement to reduced dimensions \cite{takahata,qwell,confinement}, 
and strong coupling with electromagnetic fields in an optical cavity \cite{Walther2018,microcavity1} have been investigated, paving the way for cavity QED \cite{cavityQED} and networking of superconducting qubits \cite{SC}.

In the present experiment, a high-quality natural single-crystal Cu$_2$O, 
polished into a cylinder (3 mm in diameter and height), was placed in the dielectric cavity of a flow-type cryostat and irradiated with nanosecond laser pulses.
The absorption of the microwave due to the cyclotron resonance (CR) of photoexcited carriers was measured at an angular frequency of $\omega$ ($2\pi\times$9.626 GHz) 
while sweeping the magnetic field \cite{NakaPRB} (see Supplemental Material II for experimental details \cite{SM}). 
Unlike prior studies employing CR under electron-hole pair generation across the energy gap \cite{Ohyama,NakaPSSB}, this study proposes the photoexcitation of holes utilizing deep acceptor levels attributed to copper vacancies.
This novel method suppresses the production of free electrons in the conduction band, 
thereby avoiding unfavorable changes in the energy bands caused by many-body effects, such as electron-hole interactions, bimolecular exciton collisions, and magnetoplasma formation \cite{AkimotoEPJB}.

\begin{figure}[tbp]
\centering
\includegraphics[width=8.5cm, clip]{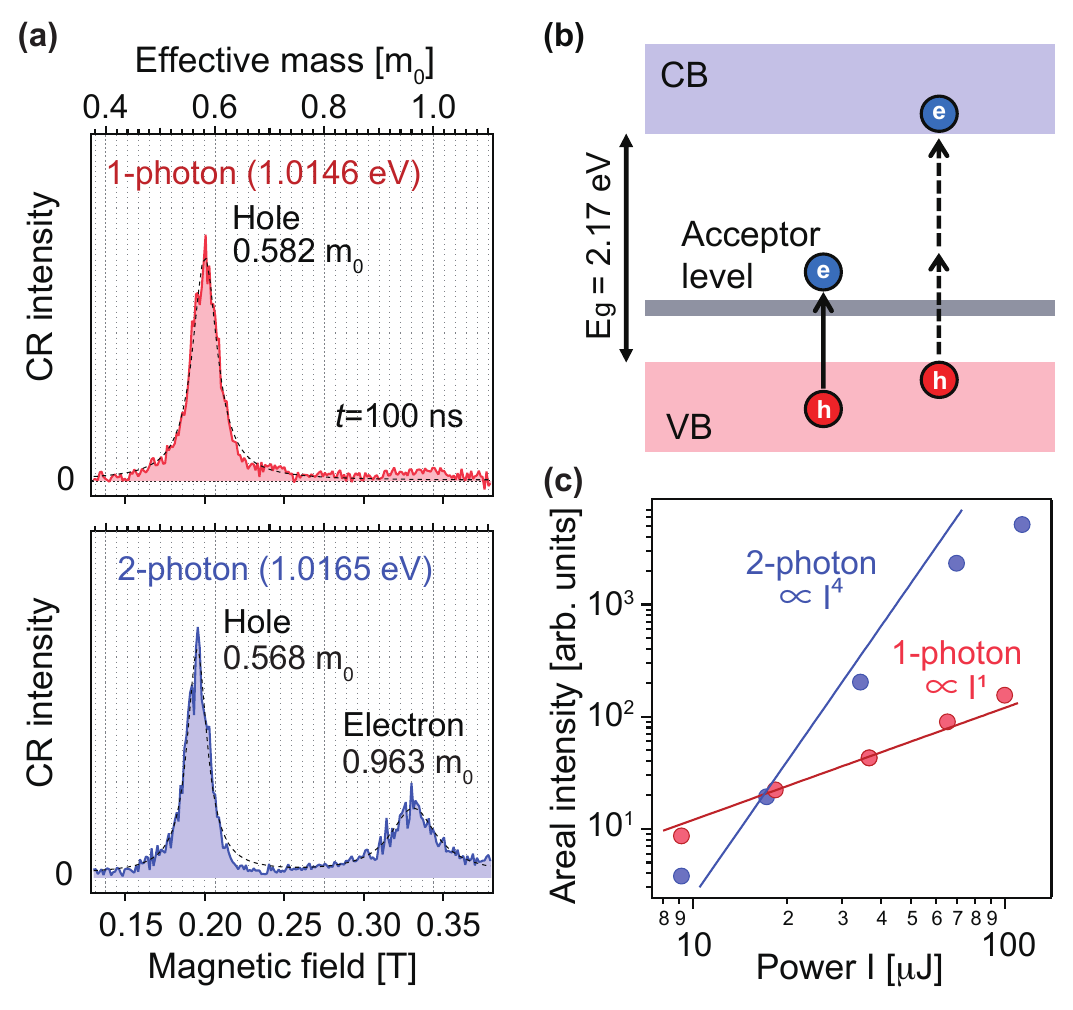}
\caption{
(a) CR spectra of Cu$_2$O at 6.1 K at a time delay of 100 ns following photoexcitation. 
Upper panel: Hole excitation by one-photon transition. Lower panel: Two-photon resonance excitation of excitons.
(b) Schematic of band structure and holes excited from the acceptor level to the valence band in Cu$_2$O.
Dashed arrows indicate the excitation scheme accompanying the generation of free electron-hole pairs.
(c) Excitation power dependence of the hole signal intensity.
}
\label{Figure2}
\end{figure}

The upper panel of Fig. 2(a) shows the CR spectrum of Cu$_2$O at 6.1 K for a time delay of $t=100$ ns. 
The laser photon energy was set at 1.0146 eV so that the two-photon energy was detuned from the yellow 1s exciton level at 2.033 eV, slightly below the energy gap. 
As illustrated by the solid arrow in Fig. 2(b), this case is expected to allow only the excitation of holes via electron transitions to the acceptor level, which is located at $\sim 0.55$ eV above the valence band maximum \cite{accexp,acctheo}.
Note that the electrons at the acceptor level are localized and do not generate CR signals. 
For reference, the lower panel in Fig. 2(a) presents a prior case in which excitons were generated via the two-photon resonant absorption of 1.0165 eV photons [see dashed arrows in Fig. 2(b)].
As the excitons are ionized into free carriers due to the two-body collisional process \cite{AkimotoEPJB}, electron resonance appears in addition to the hole resonance.

Based on the resonance magnetic field $B_0$ obtained by a Lorentzian fit (dashed line) to the CR spectrum, the effective mass
\begin{equation}
m^{\rm CR} = eB_0/\omega
\label{CR}
\end{equation}
is obtained, where $e$ is the elementary charge.
The hole mass ($0.582$ times the free electron mass $m_0$) under one-photon excitation was larger than that under two-photon excitation ($0.568 m_0$), 
which was low-field shifted owing to many-particle effects \cite{NakaPRB}. 
Figure 2(c) shows the areal intensity of the hole signals as a function of incident laser power.
Linear and quartic dependences were observed, as expected for the one-photon transition and two-body processes after the two-photon transition, respectively.

The inset of Fig. 3(a) presents a contour plot of the CR signal of the holes at 10 K, solely generated by the one-photon transition, as a function of the delay time $t$.
For quantitative analysis, the $m^{\rm CR}$ values extracted from the CR curves at each delay time are plotted by circles in Fig. 3(a).
Following photoexcitation, the holes thermally relaxed in the momentum space with increasing time [Fig. 3(b)].
If the band dispersion is parabolic, the effective mass is independent of the momentum; thus, the resonance field is independent of $t$ (dashed line). However, a clear high-field shift was observed within 50 ns. 
Notably, the direction of the initial shift was opposite to that caused by many-particle effects \cite{AkimotoEPJB}, and the steady-state values as $t \rightarrow \infty$ varied with temperature, implying the nonparabolicity of the band.
By further fitting the temporal variation using an exponential decay function, we obtained two mass values: the initial effective mass at $t=0$ and the steady-state mass as $t\rightarrow \infty$.
The best-fit curves are represented by solid lines in Fig. 3(a).
The obtained initial mass values ($\sim$0.61 $m_0$) were significantly larger than the steady-state mass values.
The validity of assuming the thermal exponential decay is detailed in Supplemental Material III \cite{SM,Takahatawide,Allen,O'Hara}.

The theoretical treatment of the hole bands requires a Hamiltonian that accounts for the coupling between the hole spin and pseudospin and describes the nonspherical symmetry of the crystal lattice \cite{Schone2016,Coupled,Scaling,Ertl2022,Impact,Even,Efield}. 
Although the Hamiltonian does not resemble that for Dirac particles in terms of the lack of $k$-linear terms owing to the inversion symmetry of Cu$_2$O (see Eq. (S3) \cite{SM}), the hole energy dispersion is well approximated by
\begin{eqnarray}
E(p)&=&\sqrt{\tilde{m}^{2}\tilde{c}^{4}+p^2 \tilde{c}^{2}}-\tilde{m}\tilde{c}^{2}
\label{dispersion}
\end{eqnarray}
for the momentum range of interest.
We may attribute the hole effective mass $m^{\rm CR}$ to the relativistic mass \cite{Landau}
\begin{eqnarray}
m^{\rm CR}(p)
=\sqrt{\tilde{m}^2+p^2/\tilde{c}^2}.
\label{relativisticmass}
\end{eqnarray}
Thus, both dispersion and mass enhancement are characterized by only two parameters, $\tilde{m}$ and $\tilde{c}$, in place of $m$ and $c$ for a relativistic particle in vacuum, or $\tilde{m}$ and $E_{\rm g}$ in the condensed matter version, Eq. (1).

\begin{figure}[tb]
\centering
\includegraphics[width=8.5cm, clip]{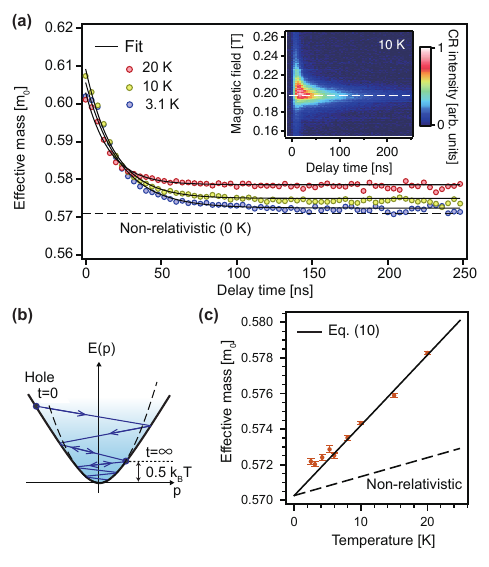}
\caption{
(a) Temporal variation of hole effective mass at $T=3.1,$ 10, and 20 K with $\sim60$ $\mu$J incidence. 
The solid lines represent best-fit exponential functions.
Inset: Contour map of the time-resolved spectra at 10 K.
(b) Schematic of hole relaxation in momentum space. The band is shown inverted relative to Fig. 1, with the vertical axis representing hole energy rather than electron energy.
(c) Temperature dependence of the hole effective mass as $t\rightarrow \infty$ (dots) with the fit using Eq. \eqref{mass_infty} (solid line).
Dashed lines in (a)$-$(c) represent the non-relativistic case of parabolic dispersion.
}
\label{Figure3}
\end{figure}

By eliminating $p$ from Eqs. (\ref{dispersion}) and (\ref{relativisticmass}), one finds 
\begin{eqnarray}
m^{\rm CR}(p)&=& \tilde{m} + E(p)/\tilde{c}^{2}. 
\label{mass_k} 
\end{eqnarray}
Now, we assume $E(p)=
\frac{1}{2} k_{\rm B}T$ at the peak of the Boltzmann distribution weighted by a three-dimensional density of states.
Here, $k_{\rm B}$ is the Boltzmann constant, and $T$ is the temperature. 
Then, Eq. \eqref{mass_k} transforms from a momentum to temperature dependence
\begin{eqnarray}
m^{\rm CR}(T)  =  \tilde{m} +\frac{k_{\rm B}T}{2\tilde{c}^2} = \tilde{m^*}\,\eta(T) +\frac{\eta(T)^2}{2\tilde{c^*}^2} k_{\rm B}T, 
\label{mCRp}
\end{eqnarray}
where 
\begin{equation}
\tilde{m}=\tilde{m^*}\,\eta(T),\, \tilde{c}=\tilde{c^*}/\eta(T)
\end{equation}
are introduced to include the polaron correction.
Note that the CR effective mass obtained using Eq. \eqref{CR} is the polaron mass, which is renormalized owing to the coupling with phonons \cite{polaron}.
The temperature-dependent correction factor is approximately expressed as
\begin{equation}
\eta(T) =1+\beta k_{\rm B} T 
\label{polaron}
\end{equation}
with $\beta=2.163$ eV$^{-1}$ \cite{note2,dielectric,polaron,Eph}. 
By substituting Eq. \eqref{polaron} into Eq. \eqref{mCRp}, we obtain
\begin{eqnarray}
m^{\rm CR}(T)=\tilde{m^*}+\left(\tilde{m^*}\beta+\frac{1}{2\tilde{c^*}^2}\right)k_{\rm B}T  \nonumber \\
+\frac{\beta}{\tilde{c^*}^2}(k_{\rm B}T)^2+\frac{\beta^2}{2\tilde{c^*}^2}(k_{\rm B}T)^3.
\label{mass_infty}
\end{eqnarray}

We applied the derived equations to the experimental data.
The steady-state mass values are plotted in Fig. 3(c) as a function of $T$.
The approximately linear $T$-dependence indicates that the higher-order terms of $T$ have only minor contributions.
By fitting the data above 6 K \cite{note3,Konishi} to Eq. (\ref{mass_infty}), we obtain
\begin{equation}
\tilde{m^*}
=(0.57027\pm 0.00014) m_0 
\end{equation}
and 
\begin{eqnarray}
\tilde{c^*}
= (1.628 \pm 0.029) \times 10^{5} \,  {\rm m/s}.
\end{eqnarray}
The fitting result is shown as a solid line. 

The physical meaning is as follows: $\tilde{m^*}\, \eta(T)=\tilde{m}$ is the polaron mass at rest (indicated by the dashed line), and $\tilde{c^*}/\eta(T)=\tilde{c}$ is the rescaled light speed. 
Substituting the $\tilde{m}$ and $\tilde{c}$ values into Eq. \eqref{dispersion} provides a nonparabolic energy dispersion.
As detailed in Supplemental Material IV \cite{SM}, the experimentally determined hole dispersion relation (for wavenumbers up to 4.1 $\%$ of the BZ) is consistent with the first-principles band calculations \cite{French}; 
the mass value increases with momentum (i.e., the band is flattened), and the enhancement is of the correct order of magnitude if the measurements are assumed to reflect the angle-averaged anisotropic band.

To date, the direct observation of the mass enhancement due to the relativity of particles in free space has not been straightforward because of the minuscule size of the nonparabolicity parameter: 
$(4m_0 c^2)^{-1} = 5\times 10^{-7}$ eV$^{-1}$ for positronium, for example.
In condensed matter, the nonparabolicity parameter is typically much larger, on the order of ${E_{\rm g}}^{-1}$ \cite{GaAs, PV}. 
However, our results for Cu$_2$O demonstrate an even higher nonparabolicity, scaled by $(2 \tilde{m^*}\tilde{c^*}^2)^{-1}=5.7$ eV$^{-1}$ $\sim |\Delta|^{-1}$.
This enhancement can be understood as the ``mimicking of an NGS" in Cu$_2$O, where the spin-orbit interaction takes over the role of the electric-dipole interaction across $E_{\rm g}$ in conventional materials (see Fig. 1) \cite{topo,Schone2016}.

To further demonstrate the benefits of determining the band dispersion, we examined the exciton binding energies.
This approach opens up a new arena for exploring relativistic bound states in condensed matter, which are unstable in NGS and have previously been observed only under extreme conditions \cite{vac}, 
beyond the reach of tabletop experiments.
Figure 4(a) depicts the absorption spectrum measured at $T=2$ K, 
using spectrally filtered light (photon energies: $2.148-2.178$ eV)
from a tungsten lamp as the probe in the low-density regime. 
The spectrum displays odd-parity Rydberg exciton peaks with different $n$ and $\ell$ values \cite{highang}.
Their energy deficits relative to the energy gap $E_{\rm g}$ (dash-dotted line) indicate the binding energies ($<0$), 
which are well reproduced by calculations based on the Dirac equation \cite{fieldtheory,energycalc} as described below. 

\begin{figure}[tbp]
\centering
\includegraphics[width=8.5cm, clip]{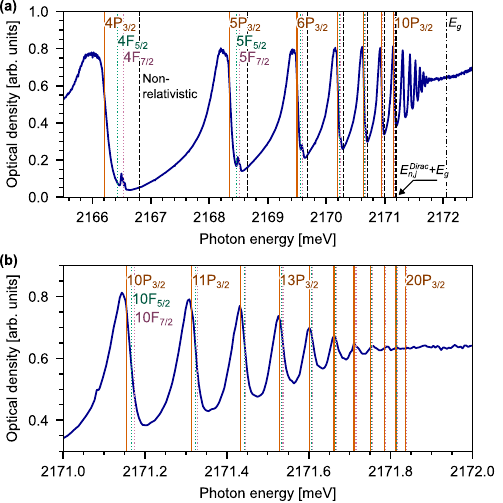}
\caption{
(a) Absorption spectrum measured at 2 K and resonance energies calculated using Eq. \eqref{binding} and $E_{\rm g}=2.17206$ eV for $n=4-10$ and $\ell=1, 3$ states (full and dotted lines) and for the non-relativistic case (dashed lines).
Peaks are saturated because of the large sample thickness. 
(b) Enlarged spectrum for higher states ($n\geq 10$).
}
\label{Figure4}
\end{figure}

Here, the two-body problem for the exciton, composed of an electron (in the parabolic CB with effective mass $m_{\rm e}$) and a hole in the nonparabolic VB1,
is transformed into a one-body problem with reduced mass $\mu(p)=1/\left( 1/m_{\rm e} +1/m^{\rm CR}(p)\right)$.
We address the eigenvalue problem of a Dirac equation for a relativistic quasiparticle rather than the non-relativistic Schr\"{o}dinger equation.
Our approach extends the Dirac theory by incorporating band nonparabolicity through the momentum dependence of $m^{\rm CR}$.
This reformulation necessitates replacing $m$, $c$, and the fine-structure constant $\alpha= e^2/(4\pi\epsilon_0 \hbar c)$ with
$\tilde{\mu}=1/(1/m_{\rm e}+1/\tilde{m})$, $\tilde{c_{\mu}}=(\tilde{m}/\tilde{\mu})^{3/2}\tilde{c}$, and
$\tilde{\alpha}= e^2/(4\pi\epsilon_{\rm b} \epsilon_0 \hbar \tilde{c_{\mu}})$, where
$\hbar$ is the reduced Planck constant and $\epsilon_0$ and $\epsilon_{\rm b}\epsilon_0$$=7.5\epsilon_0$ \cite{dielectric} are the vacuum and material permittivities, 
respectively (see Supplemental Material V for derivations \cite{SM}).
We assume $m_{\rm e}=0.983 m_0$ for zero temperature, as extrapolated from measurements at 10 K \cite{NakaPRB}.
The binding energies were then calculated as 
\begin{equation}
E_{n,j}^{\rm Dirac}= \frac{\tilde{\mu}\tilde{c_{\mu}}^2}{
\sqrt{1+ \frac{\tilde{\alpha}^2}{\left( n-(j+1/2)+\sqrt{(j+1/2)^2-\tilde{\alpha}^2}\right)^2}}}-\tilde{\mu}\tilde{c_{\mu}}^2,
\label{binding}
\end{equation} 
with total angular momentum $j=\ell\pm1/2$. 
By incorporating $E_{\rm g}$  at 2.17206 eV, 
the full and dotted lines denote the calculated resonance energies, labeled $n$P$_j$ ($\ell=1$) and $n$F$_j$ ($\ell=3$) in Figs. \ref{Figure4}(a) and \ref{Figure4}(b).
These energies accurately reproduced the observed positions (see Tables S2 and S3 \cite{SM,para,Rogers,Stolz, Stolz2022}).

In this framework, the parabolic approximation with $\tilde{\alpha}\ll1$ leads to $E_{n,j}=-\tilde{\mu} \tilde{c_{\mu}}^2\tilde{\alpha}^2/(2n^2)$ of the non-relativistic hydrogenic series [dashed lines in Fig. 4(a)].
Remarkably, a larger deviation from this case was found for P-excitons than for F-excitons. 
The quantum defect $\delta_{n,\ell}$ in Eq. (2) indexes these deviations and approximately matches $\delta_j^{\rm Dirac}=(j+1/2)-\sqrt{(j+1/2)^2-\tilde{\alpha}^2}$ in Eq. (13) \cite{Jacobs}.
Our parameters provide a Rydberg energy of $R=\tilde{\mu} \tilde{c_{\mu}}^2\tilde{\alpha}^2/2=87.3$ meV and
$\delta_{3/2}^{\rm Dirac} = 0.215$.
These values are close to $R = 87.46198$ meV and $\delta_{{\rm P}}= 0.206$ by a recent scrutiny of the experimental spectra \cite{Stolz2022,Stolz}.
Similar estimates for F-excitons ($\delta_{5/2}^{\rm Dirac} = 0.139$ and $\delta_{7/2}^{\rm Dirac} = 0.103$)
are also consistent with previously reported values \cite{Krueger, Schone2016, Scaling, Rogers} (Supplemental Material V \cite{SM}).
Our results indicate that nonparabolicity is essential for the quantum defect and that the compact expression for 
band dispersion with only two parameters effectively encapsulates the key features of Rydberg excitons.

In conclusion, by avoiding many-body effects and carefully analyzing time-resolved cyclotron resonance signals,
we experimentally quantified the valence-band dispersion of Cu$_2$O.
We demonstrated that the proposed relativistic dispersion relation, Eq. \eqref{dispersion}, 
and the extremely simple analytical eigenvalues, Eq. \eqref{binding}, 
can describe the Rydberg spectrum of excitons without delicate theoretical treatments \cite{Schone2016,Coupled,Scaling,Krueger, Ertl2026,Ertl2022,Impact,Even,Efield}. 
The detailed mechanism of hole creation is left for future studies, 
which could shed light on  environmental charge control \cite{purification}.
This study is the first to demonstrate the direct relevance between the {\it measured} band nonparabolicity and quantum defect values in solid materials. 
We believe that our findings will stimulate forthcoming theoretical and experimental investigations of the interesting properties of Rydberg states under various environmental conditions.

\begin{acknowledgments}
{\it Acknowledgments---}
We thank Kazuki Konishi (Kyoto University) for his experimental assistance and 
Hideto Matsuoka (Hokkaido University of Education, Osaka Metropolitan University) for the cryogenic technique.
This work was supported by JSPS KAKENHI Grant Numbers JP17H02910, JP22H01156/JP23K22427, JP21K03420, JP19K21849,
JSPS Bilateral Program Number JPJSBP120269924, ISHIZUE 2026 of Kyoto University, and the RIKEN TRIP initiative.
\end{acknowledgments}


\clearpage

\onecolumngrid
\begin{center}
{\large\bfseries Supplemental Material\\ Giant Rydberg Excitons Delineated as Relativistic Bound States \par}
\end{center}
\vspace{1.5em}
\twocolumngrid

\setcounter{equation}{0}
\setcounter{figure}{0}
\setcounter{table}{0}
\setcounter{page}{1}
\setcounter{section}{0}
\renewcommand{\theequation}{S\arabic{equation}}
\renewcommand{\thefigure}{S\arabic{figure}}
\renewcommand{\thetable}{S\arabic{table}}
\renewcommand{\thepage}{S\arabic{page}}
\renewcommand{\thesection}{S\Roman{section}}

\section{I. Comparison to the Kane model for narrow-gap semiconductors}

Luttinger and Kohn developed the ${\bm k} \cdot {\bm p}$ method, which describes the valence band structure of semiconductors using second-order perturbation theory.
Kane first indicated an alternative approach suitable for narrow-gap semiconductors (NGS), 
which neglects distant bands and simply treats conduction and valence bands without spin \cite{Kane}.
The Hamiltonian for this two-band model is given by \cite{Zawadzki}
\begin{equation}
\mathcal{H}=\left(
\begin{tabular}{cccc} 
$-E'+\frac{E_{\rm g}}{2}$ &$ C\hbar k_{-}$ &$C\hbar k_z$ &$C\hbar k_{+}$ \\
$C\hbar k_{+}$ &$ -E'-\frac{E_{\rm g}}{2}$&0&0 \\
$C\hbar k_{z}^{\dagger}$ &0&$ -E'-\frac{E_{\rm g}}{2}$ &0 \\
$C\hbar k_{-}$ &0&0&$-E'-\frac{E_{\rm g}}{2}$ \\
\end{tabular}
\right),
\end{equation}
where $E'=E-\hbar^2k^2/(2m_0)$, $C=-i\bra{S}p_x\ket{X}/m_0$, $k_{\pm}=(k_x\pm i k_y)/\sqrt{2}$, and $k^2=k_x^2+k_y^2+k_z^2$.
$\ket{S}$ and $\ket{X}$ represent the electron and hole wave functions, respectively, and $p_x=\hbar k_x$ is the momentum operator.
Two of the energy roots of the above matrix are expressed as
\begin{eqnarray}
E'_{1,2}(k)=\pm \sqrt{\left( \frac{E_{\rm g}}{2} \right)^2+C^2\hbar^2 k^2}  \nonumber\\
=\pm \sqrt{\left( \frac{E_{\rm g}}{2} \right)^2+p^2\frac{E_{\rm g}}{2m}}, 
\label{eqS2}
\end{eqnarray}
where $C^2=E_{\rm g}/(2m)$.
The $+$ and $-$ signs correspond to the conduction and light-hole bands, respectively.
The free-electron term $-\hbar^2k^2/(2m_0)$ in $E'$ is usually small in NGS (because of $m \ll m_0$), 
and the energy dispersion $E=E'_{1,2}(k)+\hbar^2k^2/(2m_0)\simeq E'_{1,2}(k)$ becomes analogous to Eq. (1) in the main text after subtracting an energy offset $\pm E_{\rm g}/2$ from Eq. \eqref{eqS2}.

Cuprous oxide (Cu$_2$O) is not an NGS but a medium-gap semiconductor with $E_{\rm g}=2.17$ eV. 
Because the condition $m \ll m_0$ is not satisfied, Kane's model is inapplicable to Cu$_2$O.
Furthermore, Cu$_2$O possesses an inversion symmetry that conserves parity. 
The matrix element containing $C$ is zero in Cu$_2$O because both the conduction and valence bands have even parity, and the electric dipole operator of odd parity cannot connect them.
This is evidenced by the electron effective mass of approximately $m_0$, which was also measured in the present study. 
Thus, {\it the conduction band in Cu$_2$O can be treated as a distant band when considering the valence bands.}

Unlike other cubic semiconductors, the top valence band (VB1) 
of Cu$_2$O has a split-off character because of the negative spin-orbit interaction energy.
The spin-orbit coupling of strength $\bar{\Delta}=-\Delta=131$ meV enters the Suzuki-Hensel Hamiltonian for the valence bands of Cu$_2$O \cite{Schone2016}:
\begin{eqnarray}
&&\mathcal{H}=-\frac{\bar{\Delta}}{3}({\bm I} \cdot {\bm \sigma})+\frac{\hbar^2}{2 m_0}\left\{ \vphantom{\frac{1}{3}} [A_1+B_1({\bm I}\cdot {\bm \sigma})]k^2+\right.\nonumber\\
&&\left[ A_2(I_x^2-\frac{1}{3}I^2) \right. +\left.B_2(I_x\sigma_x-\frac{1}{3}{\bm I}\cdot {\bm \sigma}) \right] k_x^2 + {\rm c.p.} +\nonumber\\
&&\left. [A_3(I_xI_y+I_yI_x)+ B_3(I_x\sigma_y+I_y\sigma_x)]\{k_x,k_y\}+ {\rm c.p.} \vphantom{\frac{1}{3}}\right\}. \nonumber \\
\end{eqnarray}
Here, ${\bm I}$ denotes the angular momentum matrices for $I=1$, ${\bm \sigma}$ the Pauli spin matrices,
$\{k_x,k_y\}=(k_xk_y+k_yk_x)/2$, and 
$A_1$, $A_2$, $A_3$, $B_1$, $B_2$, $B_3$ are dimensionless parameters determined from the band calculations \cite{French}.
Note that the Hamiltonian does not include a $k$-linear term as in Eq. (S1). 
The split-off hole energy is derived as (see Eq. (2.22a) and the definitions of $P_1$, $P_2$, $P_3$ in Ref. \cite{Thesis}.)
\begin{equation}
E_{\rm VB1}(k)=\frac{1}{12}\left (P_1k^2-6\bar{\Delta} +\sqrt{P_2k^4+P_3\bar{\Delta} k^2+36\bar{\Delta}^2} \, \right).
\label{E7}
\end{equation}
Removal of the square root by squaring Eq. \eqref{E7} leads to the relation:
\begin{eqnarray}
E_{\rm VB1}(k)^2&-&\frac{(P_1k^2-6\bar{\Delta})E_{\rm VB1}(k)}{6} +\frac{(P_1 k^2-12\bar{\Delta})P_1 k^2}{144}\nonumber\\
&=&\frac{(P_2k^2+P_3\bar{\Delta}) k^2}{144} .
\end{eqnarray}
For the wavenumber range of interest ($k<0.1$ nm$^{-1}$), 
the terms proportional to $k^2$ inside each bracket are much smaller than the other terms containing $\bar{\Delta}$.
Solving Eq. (S5) by dropping these small terms yields
\begin{eqnarray}
E_{\rm VB1}(k)\simeq -\frac{\bar{\Delta}}{2} +\sqrt{\left( \frac{\bar{\Delta}}{2} \right)^2+k^2 \frac{(12P_1+P_3)\bar{\Delta}}{144}},  
\end{eqnarray}
which reads 
\begin{equation}
E_{\rm VB1}(p)\simeq\sqrt{\left(\frac{\bar{\Delta}}{2}\right)^2+p^2 \frac{\bar{\Delta}}{2\tilde{m}_{\rm h}}}-\frac{\bar{\Delta}}{2},
\label{delta}
\end{equation}
if we express $(12P_1+P_3)/144=\hbar^2A_1/(2m_0)$ by $\hbar^2/(2\tilde{m}_{\rm h})$.
Note that $A_1$ and $\tilde{m}_{\rm h}$ are negative; thus, $E(p)$ represents an upward convex hole band. 
Eq. \eqref{delta} formally resembles Eq. (1) under correspondence of $\bar{\Delta} \Leftrightarrow E_{\rm g} $ and $\tilde{m}_{\rm h} \Leftrightarrow \tilde{m}$. 

This situation is understood more simply by learning from Ref. \cite{Zawadzki} that
"The first nonparabolic approximation to any $E(k)$ dependence is always the relativistic type."
If one expands $k^2$ to the series:
\begin{equation}
k^2(E)=a_0+a_1 E+a_2 E^2+a_3 E^3 +\cdots,
\end{equation}
and takes up to the $a_2$ term, a dispersion relation analogous to Eq. \eqref{delta} is obtained.
The effective mass approach corresponds to $a_0=a_2=0$ and $a_1\neq0$,
whereas the relativistic case corresponds to linear dispersion with $a_0=a_1=0$ and $a_2\neq0$.

These consequences validate the use of Eq. (4) in the analysis of the present data, despite the fact that Cu$_2$O is {\it not} an NGS.
In Cu$_2$O, {\it because of the unique band symmetries}, coupling to the conduction band across $E_{\rm g}$ through the electric-dipole interaction is {\it forbidden}, 
and the role is taken over by the spin-orbit interaction with the light- and heavy-hole bands across $\Delta$.

\section{II. Experimental details}

\begin{figure}[b]
\centering
\includegraphics[width=7.5cm, clip]{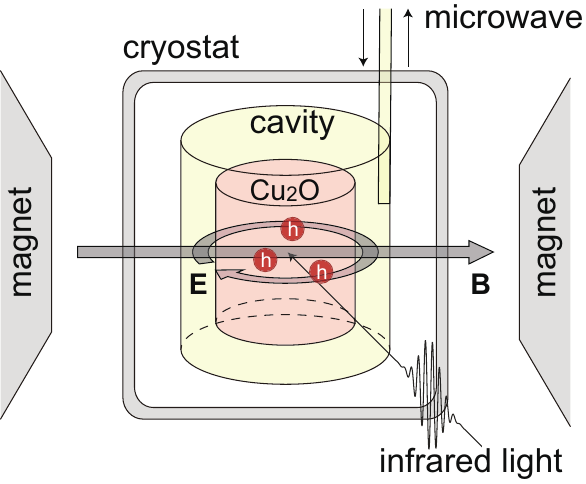}
\caption{Schematic of the setup for time-resolved cyclotron resonance measurements using a dielectric cavity.}
\label{FigureS1}
\end{figure}

Figure \ref{FigureS1} shows a schematic of the experimental setup. 
For the cyclotron resonance measurements, a high-quality natural single crystal of Cu$_2$O shaped into a cylinder with the cylinder axis oriented along the [110] crystal axis was used.
A cylindrical shape rather than a cubic shape was chosen to improve the coupling of the photoexcited holes with the microwave electric field, which is localized near the inner wall of the cylindrical dielectric cavity (Bruker, MD5W1).
A magnetic field was applied along the [001] crystal axis, which was perpendicular to the cylinder axis.
The cavity and sample were inserted into a flow-type cryostat and kept at temperatures down to 2.5 K. 
The sample was in direct contact with the helium gas flow, which served as a suitable heat sink above 6 K.
However, at lower temperatures, heating occurred owing to microwave absorption \cite{Konishi}.
The cavity quality factor was reduced from $Q=800$ in our previous setup \cite{AkimotoEPJB} to $Q=200$ to achieve a higher temporal resolution of 4 ns.
The microwave absorption was measured using a bridge (Bruker, ELEXSYS E580).
The microwave angular frequency was recorded at each scan with an accuracy of $\pm$1 kHz.

\begin{figure}[t]
\centering
\includegraphics[width=8.5cm, clip]{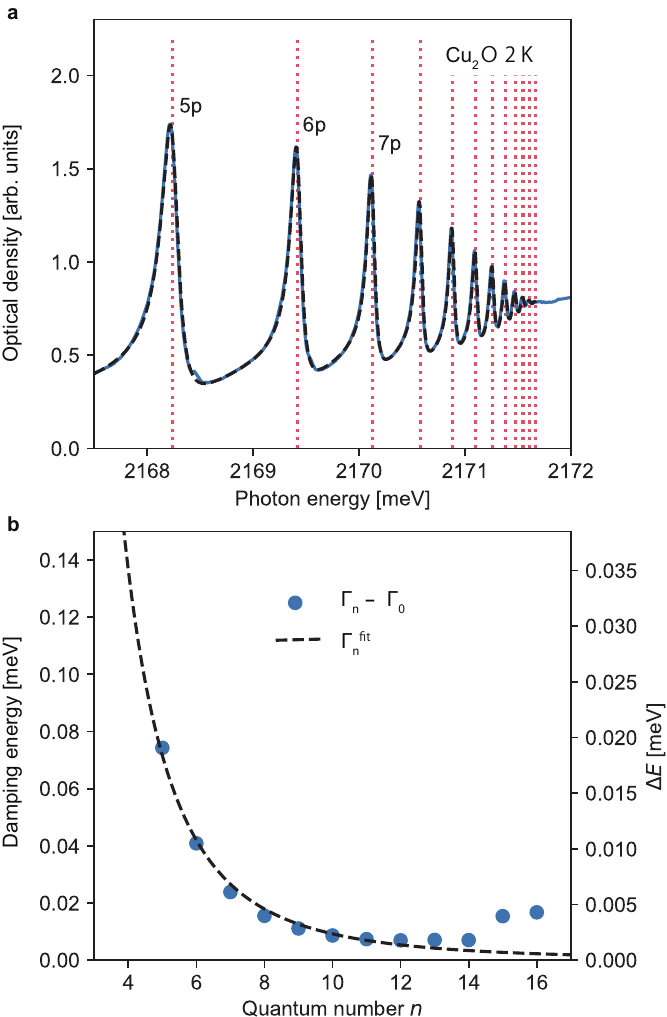} 
\caption{
(a) Absorption spectrum of Sample 1 at 2 K. 
The dashed line represents the fit by $\sum_n\alpha_n(E)$. 
The dotted lines indicate positions of resonance energies. 
(b) Homogeneous damping energy $\Gamma_n-\Gamma_0$ (solid dots) and the best fit $\Gamma_n^{\rm fit}$ (dashed line) as a function of $n$, where $\Gamma_0=0.02$ meV.
}
\label{FigureS2}
\end{figure}

\begin{table*}[bthp]
\caption{
Resonance energy $E_n$, damping energy $\Gamma_n$, and positions $E_{\rm max}$  of the maximum peak height in (a) Sample 1 and (b) Sample 2. 
Resonance energies in Sample 2 were estimated by considering $\Delta E$ based on the theoretical spectral shape with a homogeneous damping energy $\Gamma_n^{\rm fit}$ (see text for details).
The bandgap energies $E_{\rm g}=2172.008\underline{2}$ meV and 2172.062\underline{2} meV were determined by the fit of $E_n$ to Eq. (2), respectively, for Sample 1 and Sample 2.  
All values, including binding energies $E_{n,\ell}=-E_{\rm g}+E_n$, are given in units of meV. 
}
\tabcolsep=10pt
\begin{tabular}{|c||c|c|c||c|c|c|c|c|}\hline 
\multicolumn{1}{|c||}{} &\multicolumn{3}{|c||}{(a) Sample 1} &\multicolumn{5}{|c|}{(b) Sample 2}\\  \hline
$n$ &  $E_n$ & $\Gamma_n$ & $E_{n,\ell}$ & $E_{\rm max}$ & $\Gamma_n^{\rm fit}$ & $\Delta E$ & $E_n=E_{\rm max}+\Delta E$ & $E_{n,\ell}$ \\ \hline  \hline 
2 & $-$ & $-$ & $-$ & $-$ & $0.87375$ & $0.22443$ & $-$ & $-$ \\ \hline
3 & $-$ & $-$ & $-$ & $-$ & $0.30683$ & $0.07881$ & $-$ & $-$ \\ \hline
4 & $-$ & $-$ & $-$ & $2166.022$ & $0.13652$ & $0.03507$ & $2166.057$ & $-6.005$ \\ \hline
5 & $2168.242$ & $0.09429$ & $-3.766$ & $2168.197$ & $0.07158$ & $0.01839$ & $2168.216$ & $-3.846$\\ \hline
6 & $2169.423$ & $0.06088$ & $-2.585$ & $2169.427$ & $0.04195$ & $0.01078$ & $2169.438$ & $-2.624$\\ \hline
7 & $2170.126$ & $0.04385$ & $-1.882$ & $2170.154$ & $0.02662$ & $0.00684$ & $2170.161$ & $-1.901$\\ \hline
8 & $2170.578$ & $0.03554$ & $-1.430$ & $2170.615$ & $0.01792$ & $0.00460$ & $2170.620$ & $-1.442$\\ \hline
9 & $2170.884$ & $0.03111$ & $-1.124$ & $2170.923$ & $0.01263$ & $0.00324$ & $2170.927$ & $-1.135$\\ \hline
10 & $2171.101$ & $0.02866$ & $-0.907$ & $2171.144$ & $0.00923$ & $0.00237$ & $2171.146$ & $-0.916$\\ \hline
11 & $2171.261$ & $0.02738$ & $-0.747$ & $2171.308$ & $0.00694$ & $0.00178$ & $2171.310$ & $-0.752$\\ \hline
12 & $2171.382$ & $0.02696$ & $-0.626$ & $2171.432$ & $0.00536$ & $0.00138$ & $2171.433$ & $-0.629$\\ \hline
13 & $2171.475$ & $0.02712$ & $-0.533$ & $2171.526$ & $0.00422$ & $0.00108$ & $2171.527$ & $-0.535$\\ \hline
14 & $2171.549$ & $0.02705$ & $-0.459$ & $2171.602$ & $0.00338$ & $0.00087$ & $2171.603$ & $-0.459$\\ \hline
15 & $2171.609$ & $0.03542$ & $-0.399$ & $2171.663$ & $0.00275$ & $0.00071$ & $2171.664$ & $-0.398$\\ \hline
16 & $2171.669$ & $0.03676$ & $-0.339$ & $2171.711$ & $0.00227$ & $0.00058$ & $2171.711$ & $-0.351$\\ \hline
17 & $-$ & $-$ & $-$ & $2171.758$ & $0.00189$ & $0.00049$ & $2171.758$ & $-0.304$ \\ \hline
18 & $-$ & $-$ & $-$ & $2171.791$ & $0.00159$ & $0.00041$ & $2171.791$ & $-0.271$ \\ \hline
19 & $-$ & $-$ & $-$ & $2171.815$ & $0.00136$ & $0.00035$ & $2171.815$ & $-0.247$ \\ \hline 
$E_{\rm g}$ & $2172.008\underline{2}$ & $-$ & $0$ & $-$ & $-$ & $-$ & $2172.062\underline{2}$ & $0$ \\ \hline 
\end{tabular}
\label{TableS1}
\end{table*}

The excitation light source was an optical parametric oscillator (Spectra Physics, MOPO) pumped by an Nd:YAG laser operating at a repetition rate of 10 Hz.
The infrared light with a pulse duration of 5 ns and at a wavelength of $\sim$1.1 $\mu$m was almost transparent through Cu$_2$O, 
as the photon energy was well below the energy gap. The laser beam was incident along the [$1\overline{1}0$] crystal axis of the sample. 
A low laser power of $\sim60$ $\mu$J/pulse with defocused illumination in an area of approximately 10 mm$^2$ was employed.
Hence, a homogeneous and dilute distribution of photoexcited holes was achieved throughout the entire sample volume. 
This is contrary to the conventional excitation method that creates high-density electron-hole pairs near the surface using visible light \cite{Ohyama}. 
Owing to the high detection sensitivity brought by the optimized sample shape, signals could be probed even at this low carrier density. 

For the absorption measurements,
two thin Cu$_2$O slabs were cut from the same mother crystal as for the cyclotron resonance. 
After mechanical polishing of the surfaces, the thicknesses were measured as 59 $\mu$m (Sample 1) and 210 $\mu$m (Sample 2). 
White light from a halogen lamp was filtered into the wavelength range of 569$-$577 nm (Thorlabs MF565-24 and Sigmakoki RSF-25C-561RU) and focused onto the sample mounted in an immersion-type cryostat. 
The temperature was maintained at 2 K using superfluid helium. 
The light transmitted through the sample was dispersed by a monochromator with an effective focal length of 3000 mm (Jobin Yvon, THR1500, double passage) and detected by a silicon charge-coupled device camera (Andor, DU940N). 
We corrected the wavelength shift in the monochromator caused by  air temperature fluctuations during the exposure time. 
The wavelength was calibrated with a precision of $\pm$2 $\mu$eV using four spectral lines of a neon lamp. 
The logarithm of the transmission was regarded as the optical density in arbitrary units. 
A representative spectrum obtained for Sample 1 is shown in Fig. \ref{FigureS2}(a).

The spectra were analyzed using a least-squares fit to the sum of the asymmetric Lorentz functions 
\begin{equation}
\alpha_n(E)=c_n\frac{\Gamma_n/2 + A(E-E_n)}{(\Gamma_n/2)^2+(E-E_n)^2},
\label{EqS1}
\end{equation}
for the P-excitons and Gaussian functions for the F-excitons. Here, the resonance energy $E_n$, damping energy $\Gamma_n$, 
asymmetry factor $A$ (assumed to be independent of $n$), and amplitude $c_n$ were treated as adjustable parameters,
and a linear function was added as the baseline.
The best-fit function is represented by the dashed line in Fig. \ref{FigureS2}(a), with the $E_n$ values indicated by vertical lines and the $\Gamma_n$ values plotted as solid dots in Fig. \ref{FigureS2}(b). 
The parameter values are listed in Table \ref{TableS1}(a).

Fitting the absorption spectrum measured for Sample 2 is difficult because of the saturation at the P-exciton peaks owing to the large sample thickness [see Fig. 4(a) in the main text]. 
Therefore, we extracted 
the positions of the maximum absorption, $E_{\rm max}$, and calculated the shift of the resonance energy $E_n$ from $E_{\rm max}$, that is
\begin{equation}
\Delta E= E_n-E_{\rm max}= -\frac{\Gamma_n(\sqrt{1+4A^2}-1)}{2A}.
\label{EqS2}
\end{equation} 
The dashed line in Fig. \ref{FigureS2}(b) depicts the best-fit function $\Gamma_n^{\rm fit}=9.32(n^2-1)/n^5$ meV of the damping energy after removing an $n$-independent offset of $\Gamma_0=0.02$ meV,
which is considered to be inhomogeneous broadening.
By setting $A=-0.275$ \cite{Stolz2022} and $\Gamma_n=\Gamma_n^{\rm fit}$ in Eq. \eqref{EqS2}, 
we obtained the energy shift $\Delta E$ indicated on the right axis.
The resonance energies were then obtained as $E_n = E_{\rm max}+\Delta E$, assuming that Samples 1 and 2 have the same homogeneous broadening.
Note that the resonance energies are slightly higher than $E_{\rm max}$ because $A$ is negative.
Table \ref{TableS1}(b) summarizes the $E_{\rm max}$, $\Gamma_n^{\rm fit}$, $\Delta E$, and $E_n$ values obtained.

As the quantum defect is known to have an $n$-dependence for small $n$ \cite{Stolz}, we used the resonance energies of high $n$ ($n \ge 10$) states for the fit using Eq. (2) in the main text. 
By assuming an $n$-independent quantum defect,
we found an energy gap of $E_{\rm g} = 2172.008\underline{2}$ meV for Sample 1 and 2172.062\underline{2} meV for Sample 2, respectively.
The binding energies were calculated as $E_{n,\ell}=-E_{\rm g}+E_n$ and are listed in Table \ref{TableS1}.
We note that the values in the table include $\pm 2$ $\mu$eV error in determining $E_{\rm g}$ using the fit,
in addition to the $\pm 2$ $\mu$eV error in the wavelength calibration mentioned above.

\section{III. Hole relaxation}

Here, we discuss the delay-time dependence of the measured effective mass, which is governed by changes in the kinetic energy $E(t)$ of holes.
Following laser excitation, the initially high kinetic energy of the holes decreased with time owing to momentum relaxation caused by the emission of phonons [see Fig. 3(b) of the main text]. 
The holes relax by optical phonon scattering much faster than our nanosecond temporal resolution ($t_0$),
followed by a gradual relaxation by acoustic phonon scattering.
Considering the heat transfer between two objects (holes and the lattice) in contact, the temporal change is expressed as
\begin{align}
E(t) =E(\infty)+[E(t_0)-E(\infty)]\exp [-\gamma (t-t_0)],
\label{energy_time}
\end{align}
according to the two-temperature model \cite{Allen}, where $\gamma$ is the relaxation rate.
When the band dispersion is nonparabolic, the effective mass depends on $E(t)$ and thus becomes time-dependent, as discussed below. 

For the holes, an increase in time corresponds to a decrease in momentum, and the effective mass changes according to the same exponential decay function as in Eq. (\ref{energy_time}).
Therefore, we performed exponential fits with setting $E(\infty)=0$, as shown by the solid lines in Fig. 3(a). 

\begin{figure}[t]
\centering
\includegraphics[width=8.5cm, clip]{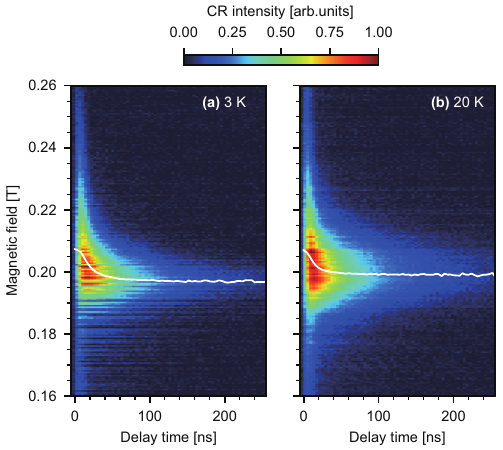}
\caption{
Contour map of the time-resolved cyclotron resonance spectra at (a) 3 K and (b) 20 K with $\sim 60$ $\mu$J incidence.
Solid lines indicate resonance magnetic fields at each time.
}
\label{FigureS3}
\end{figure}

Figure \ref{FigureS3} presents contour plots of the CR spectra obtained at (a) 3 K and (b) 20 K, respectively. 
The initial mass value obtained by the exponential fit was converted to kinetic energy using Eq. (6).
We find $E(t_0) = 4.5$ meV and $\gamma = 45$, 53, and 63 $\mu$s$^{-1}$ by the fits, respectively, at $T=3$, 10, and 20 K.
The relaxation time of $1/\gamma\sim 20$ ns is not far from the longitudinal-acoustic phonon scattering time of 2.7 ns \cite{O'Hara},
indicating that holes rapidly relax by optical-phonon scattering, followed by gradual relaxation by acoustic-phonon scattering.
The highest $E(t_0)$ appears to be limited by the energy (10.5 meV) of the $\Gamma_5^-$ optical phonon \cite{Takahatawide},
which determines the initial effective mass at $t=t_0$.

\begin{figure}[t]
\centering
\includegraphics[width=8.5cm]{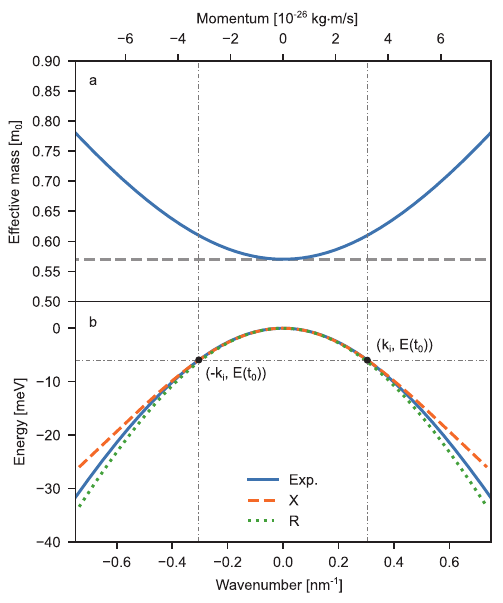} 
\caption{
(a) Wavenumber dependence of the hole effective mass around the $k=0$ ($\Gamma$) point, calculated with $\tilde{m^*}= 0.57027 m_0$, $\tilde{c^*}= 1.62755 \times 10^{5}$ m/s, and $T=0$ K.
The dashed line represents the non-relativistic case of parabolic dispersion.
(b) Solid line: valence band dispersion obtained using Eq. (4) based on our experimental results.
Dashed and dotted lines: valence band dispersions along $\Gamma-$X and $\Gamma-$R directions, respectively, from first-principles calculations \cite{French}. 
Dash-dotted lines indicate the initial wavenumber and energy.
}
\label{FigureS4}
\end{figure}

\section{IV. Band dispersion}

Using the $\tilde{m}$ and $\tilde{c}$ parameters determined in the main text, the wavenumber ($k=p/\hbar$) dependence [Eq. (5)] of the hole effective mass was obtained, as shown in Fig. \ref{FigureS4}(a) for zero temperature. 
The dashed line represents the effective mass of the parabolic band in the non-relativistic limit of  $1/\tilde{c} \rightarrow 0$.
Furthermore, the band dispersion [Eq. (4) using Eqs. (8), (9), (11), and (12)] including nonparabolicity is depicted by the solid line in Fig. \ref{FigureS4}(b). 
Note that the vertical axis represents the electron energy, whereas that in Fig. 3(b) represents the hole energy.
Dash-dotted lines indicate $E(t_0)$ and the corresponding initial wavenumber $k_{\rm i}$ at 4.1 $\%$ of the BZ,  
which was obtained from the initial effective mass value substituted into the relativistic mass expression Eq. (5).
The dashed and dotted lines represent the dispersion curves derived from the spin-dependent density functional theory (DFT) calculations along the $\Gamma-$X and $\Gamma-$R directions, respectively \cite{French}. 
The experimentally determined dispersion is between these two calculations for different wavevector directions and is considered to represent the angle-averaged band dispersion in an isotropic approximation. 
The origin of the nonparabolicity lies in the spin-orbit coupling with deeper valence bands, 
for which a mass reduction with a negative $\tilde{c}$ value is expected \cite{Coupled}, leading to a positive quantum defect value. 

\section{V. Calculation details}

In this section, we provide the calculation details.
First, we derive the eigenvalues of the excitonic bound states [Eq. (13) of the main text] based on the solution of the Dirac equation by assuming a nonparabolic valence band. 
Second, we show the approximate wavefunctions in reciprocal space for comparison with the valid wavenumber range for the determined nonparabolicity. 
Third, we discuss the quantum defect values obtained in the present and previous studies.

\begin{table*}[tb]
\caption{
Magnitude of the binding energies ($-E_{n,j}^{\rm Dirac}$) of odd-parity excitons calculated using Eq. \eqref{DiracEq_NP} and the corresponding experimental values.
$n$P$_{1/2}$ states are not shown because they are dipole-forbidden and are not expected to appear in one-photon absorption spectra.
Experimental energies of F-excitons were averaged over the fine-structure states. All values are shown in meV. 
}
\label{Diraccalc}
\tabcolsep=3pt
\begin{tabular}{|c||c|c|c|c|c||c|c|c|c||c|c|c|}\hline 
&\multicolumn{5}{c||}{P-exciton}&\multicolumn{4}{c||}{F-exciton} & \multicolumn{3}{c|}{H-exciton}\\ \hline
$n$ & \multicolumn{1}{c|}{$-E_{n,j}^{\rm Dirac}$} &Sample 1&Sample 2&Ref.\cite{Kazimierczuk2014}& Ref.\cite{Stolz}& 
\multicolumn{2}{c|}{$-E_{n,j}^{\rm Dirac}$} &Sample 2&Refs. & \multicolumn{2}{c|}{$-E_{n,j}^{\rm Dirac}$} &Refs. \\ 
 &  $n$P$_{3/2}$ &&&*Ref.\cite{Julian}& &
$n$F$_{5/2}$ & $n$F$_{7/2}$ &&\cite{highang,Krueger} & $n$H$_{9/2}$ & $n$H$_{11/2}$ &\cite{Julian,highang}\\ \hline\hline
1 & $-$ & $-$ & $-$ & $-$ & $-$ & $-$ & $-$ & $-$ & $-$ & $-$ & $-$ & $-$\\ \hline 
2 & $23.064$ & $-$ & $-$ & $23.7$ & $23.6770$ & $-$ & $-$ & $-$ & $-$ & $-$ & $-$ & $-$\\ \hline 
3 & $10.440$ & $-$ & $-$ & $10.72$ & $10.7200$ & $-$ & $-$ & $-$ & $-$ & $-$ & $-$ & $-$\\ \hline 
4 & $5.846$ & $-$ & $6.005$ & $5.98$ & $5.9860$ & $5.626$ & $5.527$ & $5.537$ & $5.5265$ & $-$ & $-$ & $-$\\ \hline 
5 & $3.714$ & $3.766$ & $3.846$ & $3.78$ & $3.7804$ & $3.601$ & $3.550$ & $3.556$ & $3.5475$ & $-$ & $-$  & $-$\\ \hline 
6 & $2.562$ & $2.585$ & $2.624$ & $2.59$ & $2.5941$ & $2.497$ & $2.467$ & $2.472$ & $2.4620$ & $2.450$ & $2.439$ & $2.433$\\ \hline 
7 & $1.871$ & $1.882$ & $1.901$ & $1.892$ & $1.8878$ & $1.831$ & $1.812$ & $1.813$ & $1.8054$ & $1.801$ & $1.794$ & $1.788$\\ \hline 
8 & $1.426$ & $1.430$ & $1.442$ & $1.440$ & $1.4366$ & $1.399$ & $1.386$ & $-$ & $1.3783$ & $1.379$ & $1.374$ & $1.372$\\ \hline 
9 & $1.122$ & $1.124$ & $1.135$ & $1.130$ & $1.1276$ & $1.103$ & $1.094$ & $-$ & $1.0785$ & $1.089$ & $1.086$ & $-$\\ \hline 
10 & $0.906$ & $0.907$ & $0.916$ & $0.911$ & $0.9081$ & $0.892$ & $0.886$ & $-$ & $0.8846$ & $0.882$ & $0.880$ & $-$\\ \hline 
11 & $0.747$ & $0.747$ & $0.752$ & $0.751$ & $0.7471$ & $0.736$ & $0.731$ & $-$ & $0.7261$ & $0.729$ & $0.727$ & $-$\\ \hline 
12 & $0.626$ & $0.626$ & $0.629$ & $0.628$ & $0.6252$ & $0.618$ & $0.614$ & $-$ & $0.6062$ & $0.612$ & $0.611$ & $-$\\ \hline 
13 & $0.532$ & $0.533$ & $0.535$ & $0.533$ & $0.5307$ & $0.526$ & $0.523$ & $-$ & $-$ & $0.521$ & $0.520$ & $-$\\ \hline 
14 & $0.458$ & $0.459$ & $0.459$ & $0.4586$ & $0.4559$ & $0.453$ & $0.451$ & $-$ & $-$ & $0.449$ & $0.448$ & $-$\\ \hline 
15 & $0.398$ & $0.399$ & $0.398$ & $0.3987$ & $0.3959$ & $0.394$ & $0.392$ & $-$ & $-$ & $0.391$ & $0.390$ & $-$\\ \hline 
16 & $0.349$ & $0.339$ & $0.351$ & $0.3497$ & $0.3460$ & $0.346$ & $0.345$ & $-$ & $-$ & $0.344$ & $0.343$ & $-$\\ \hline 
17 & $0.309$ & $-$ & $0.304$ & $0.3092$ & $0.3060$ & $0.306$ & $0.305$ & $-$ & $-$ & $0.304$ & $0.304$ & $-$\\ \hline 
18 & $0.275$ & $-$ & $0.271$ & $0.2756$ & $0.2722$ & $0.273$ & $0.272$ & $-$ & $-$ & $0.271$ & $0.271$ & $-$\\ \hline 
19 & $0.247$ & $-$ & $0.247$ & $0.2475$ & $0.2436$ & $0.245$ & $0.244$ & $-$ & $-$ & $0.243$ & $0.243$ & $-$\\ \hline 
20 & $0.223$ & $-$ & $-$ & $0.2230$ & $0.2199$ & $0.221$ & $0.220$ & $-$ & $-$ & $0.220$ & $0.219$ & $-$\\ \hline 
21 & $0.202$ & $-$ & $-$ & $0.2021$ & $0.1995$ & $0.200$ & $0.200$ & $-$ & $-$ & $0.199$ & $0.199$ & $-$\\ \hline 
22 & $0.184$ & $-$ & $-$ & $0.1839$ & $0.1827$ & $0.182$ & $0.182$ & $-$ & $-$ & $0.181$ & $0.181$ & $-$\\ \hline 
23 & $0.168$ & $-$ & $-$ & $0.1677$ & $0.1677$ & $0.167$ & $0.166$ & $-$ & $-$ & $0.166$ & $0.166$ & $-$\\ \hline 
24 & $0.154$ & $-$ & $-$ & $0.1543$ & $0.1536$ & $0.153$ & $0.153$ & $-$ & $-$ & $0.152$ & $0.152$ & $-$\\ \hline 
25 & $0.142$ & $-$ & $-$ & $0.1410$ & $0.1410$ & $0.141$ & $0.141$ & $-$ & $-$ & $0.140$ & $0.140$ & $-$\\ \hline 
26 & $0.131$ & $-$ & $-$ & $0.1305^*$ & $-$ & $0.130$ & $0.130$ & $-$ & $-$ & $0.130$ & $0.130$ & $-$\\ \hline 
27 & $0.122$ & $-$ & $-$ & $0.1205^*$ & $-$ & $0.121$ & $0.121$ & $-$ & $-$ & $0.120$ & $0.120$ & $-$\\ \hline 
28 & $0.113$ & $-$ & $-$ & $0.1095^*$ & $-$ & $0.112$ & $0.112$ & $-$ & $-$ & $0.112$ & $0.112$ & $-$\\ \hline 
29 & $0.105$ & $-$ & $-$ & $-$ & $-$ & $0.105$ & $0.104$ & $-$ & $-$ & $0.104$ & $0.104$ & $-$\\ \hline 
30 & $0.098$ & $-$ & $-$ & $-$ & $-$ & $0.098$ & $0.098$ & $-$ & $-$ & $0.097$ & $0.097$ & $-$\\ \hline 
$E_{\rm g}$&$-$&2172.008$\underline{2}$&2172.062\underline{2}&2172.0745 &2017.0701&$-$&$-$&2172.062\underline{2}&2172.0733
&$-$&$-$ &2172.08\\
&$-$&&&Ref.\cite{Kazimierczuk2014}&Ref.\cite{Stolz}&$-$&$-$&&Ref.\cite{Krueger}&$-$&$-$ &Ref.\cite{Julian}\\ \hline
\end{tabular}
\label{eps75_odd}
\end{table*}

\begin{table*}[tb]
\caption{
Magnitude of the binding energies ($-E_{n,j}^{\rm Dirac}$) for even-parity excitons calculated using Eq. \eqref{DiracEq_NP} and the corresponding experimental values from literature.
The Dirac approach does not include the spin exchange interaction acting on orthoexcitons.
Experimental energies of D-excitons were averaged over the fine-structure states. All values are shown in meV. 
}
 \label{Schcalc}
\tabcolsep=6pt
\begin{tabular}{|c||c|c|c|c||c|c|c||c|c|c|}\hline 
&\multicolumn{4}{c||}{S-exciton} &\multicolumn{3}{c||}{D-exciton} &\multicolumn{3}{c|}{G-exciton} \\ \hline
& $-E_{n,j}^{\rm Dirac}$ & \multicolumn{2}{c|}{Ref.\cite{para}} & Refs.\cite{Schone2016,Krueger} & \multicolumn{2}{c|}{$-E_{n,j}^{\rm Dirac}$} & Refs.\cite{Schone2016,Krueger} & \multicolumn{2}{c|}{$-E_{n,j}^{\rm Dirac}$} &Ref.\cite{Julian} \\ 
$n$& $n$S$_{1/2}$ & para & ortho & ortho & $n$D$_{3/2}$ & $n$D$_{5/2}$ &  & $n$G$_{7/2}$ & $n$G$_{9/2}$ & \\ \hline \hline 
1 & $121.995$ & $151.43$ & $139.31$ & $-$ & $-$ & $-$ & $-$ & $-$ & $-$ & $-$ \\ \hline 
2 & $33.045$ & $33.10$ & $34.37$ & $-$ & $-$ & $-$ & $-$ & $-$ & $-$ & $-$ \\ \hline 
3 & $13.400$ & $13.02$ & $11.63$ & $11.8033$ & $10.440$ & $9.929$ & $9.7833$ & $-$ & $-$ & $-$\\ \hline 
4 & $7.051$ & $7.05$ & $6.53$ & $6.5183$ & $5.846$ & $5.626$ &  $5.6013$ & $-$ & $-$ & $-$ \\ \hline 
5 & $4.312$ & $4.42$ & $4.14$ & $4.0493$ & $3.714$ & $3.601$ &  $3.5785$ & $3.550$ & $3.521$ & 3.523 \\ \hline 
6 & $2.900$ & $2.89$ & $2.73$ & $2.7503$ & $2.562$ & $2.497$ & $2.4966$ & $2.467$ & $2.450$ & 2.444 \\ \hline 
7 & $2.080$ & $-$ & $-$ & $1.9953$ & $1.871$ & $1.831$ & $1.8093$& $1.812$ & $1.801$ & 1.795\\ \hline 
8 & $1.564$ & $-$ & $-$ & $1.5033$ & $1.426$ & $1.399$ & $1.3800$ & $1.386$ & $1.379$ & 1.375\\ \hline 
9 & $1.218$ & $-$ & $-$ & $1.1743$ & $1.122$ & $1.103$ & $-$ & $1.094$ & $1.089$ & $-$\\ \hline 
10 & $0.975$ & $-$ & $-$ & $0.9343$ & $0.906$ & $0.892$ & $-$ & $0.886$ & $0.882$ & $-$\\ \hline 
11 & $0.798$ & $-$ & $-$ & $0.7824$ & $0.747$ & $0.736$ & $-$ & $0.731$ & $0.729$ & $-$\\ \hline 
12 & $0.665$ & $-$ & $-$ & $0.6573$ & $0.626$ & $0.618$ & $-$ & $0.614$ & $0.612$ & $-$\\ \hline 
13 & $0.563$ & $-$ & $-$ & $0.5563$ & $0.532$ & $0.526$ & $-$ & $0.523$ & $0.521$ & $-$\\ \hline 
14 & $0.482$ & $-$ & $-$ & $-$ & $0.458$ & $0.453$ & $-$ & $0.451$ & $0.449$ & $-$\\ \hline 
15 & $0.418$ & $-$ & $-$ & $-$ & $0.398$ & $0.394$ & $-$ & $0.392$ & $0.391$ & $-$\\ \hline 
16 & $0.366$ & $-$ & $-$ & $-$ & $0.349$ & $0.346$ & $-$ & $0.345$ & $0.344$ & $-$\\ \hline 
17 & $0.323$ & $-$ & $-$ & $-$ & $0.309$ & $0.306$ & $-$ & $0.305$ & $0.304$ & $-$\\ \hline 
18 & $0.287$ & $-$ & $-$ & $-$ & $0.275$ & $0.273$ & $-$ & $0.272$ & $0.271$ & $-$\\ \hline 
19 & $0.257$ & $-$ & $-$ & $-$ & $0.247$ & $0.245$ & $-$ & $0.244$ & $0.243$ & $-$\\ \hline 
20 & $0.231$ & $-$ & $-$ & $-$ & $0.223$ & $0.221$ & $-$ & $0.220$ & $0.220$ & $-$\\ \hline 
21 & $0.209$ & $-$ & $-$ & $-$ & $0.202$ & $0.200$ & $-$ & $0.200$ & $0.199$ & $-$\\ \hline 
22 & $0.190$ & $-$ & $-$ & $-$ & $0.184$ & $0.182$ & $-$ & $0.182$ & $0.181$ & $-$\\ \hline 
23 & $0.173$ & $-$ & $-$ & $-$ & $0.168$ & $0.167$ & $-$ & $0.166$ & $0.166$ & $-$\\ \hline 
24 & $0.159$ & $-$ & $-$ & $-$ & $0.154$ & $0.153$ & $-$ & $0.153$ & $0.152$ & $-$\\ \hline 
25 & $0.146$ & $-$ & $-$ & $-$ & $0.142$ & $0.141$ & $-$ & $0.141$ & $0.140$ & $-$\\ \hline 
26 & $0.135$ & $-$ & $-$ & $-$ & $0.131$ & $0.130$ & $-$ & $0.130$ & $0.130$ & $-$\\ \hline 
27 & $0.125$ & $-$ & $-$ & $-$ & $0.122$ & $0.121$ & $-$ & $0.121$ & $0.120$ & $-$\\ \hline 
28 & $0.116$ & $-$ & $-$ & $-$ & $0.113$ & $0.112$ & $-$ & $0.112$ & $0.112$ & $-$\\ \hline 
29 & $0.108$ & $-$ & $-$ & $-$ & $0.105$ & $0.105$ & $-$ & $0.104$ & $0.104$ & $-$\\ \hline 
30 & $0.101$ & $-$ & $-$ & $-$ & $0.098$ & $0.098$ & $-$ & $0.098$ & $0.097$ & $-$\\ \hline 
$E_{\rm g}$ & $-$ & 2172.08 & 2172.08 & 2172.0733 & $-$ & $-$ & 2172.0733& $-$ & $-$ &2172.08\\ 
& $-$ & Ref.\cite{Julian}&Ref.\cite{Julian}&Ref.\cite{Krueger} & $-$ & $-$ & Ref.\cite{Krueger}& $-$ & $-$ & Ref.\cite{Julian} \\ \hline
\end{tabular}
\label{oddparity}
\end{table*}

\subsection{Dirac approach}
The two-body problem for an exciton composed of an electron and a hole can be reduced to a one-body problem characterized by a reduced mass. 
We consider the nonparabolic valence band expressed using $\tilde{m}$ and $\tilde{c}$ for the hole and a parabolic conduction band with a (constant) electron effective mass. 
Owing to the fact that the split-off hole band has an angular momentum of $1/2$ (similar to the proton spin, instead of 3/2 for light and heavy holes),
the Dirac's relativistic energy spectrum for a hydrogen atom, consisting of a proton and an electron, is applicable to the present case.
Here, we assume that the center-of-mass motion of the exciton is sufficiently slow at cryogenic temperatures and that the corresponding momentum can be approximated to be zero. 
In this case, the momentum $p$ of the translational motion of the hole is equal to the relative momentum of the exciton.
In analogy with Eq. (5), we wish to express the momentum ($p$) dependence of the reduced mass as
\begin{equation}
\mu(p)=\sqrt{\tilde{\mu}^2+p^2/\tilde{c_{\mu}}^2},
\label{mup}
\end{equation} 
where the expressions of $\tilde{\mu}$ and $\tilde{c_{\mu}}$ are to be determined below.

Based on the definition of $\mu(p)$, one obtains
\begin{eqnarray}
\frac{1}{\mu(p)}&=&\frac{1}{m_{\rm e}}+\frac{1}{m^{\rm CR}(p)}=\frac{1}{m_{\rm e}}+\frac{1}{\tilde{m} 
\sqrt{1+p^2/(\tilde{m} \tilde{c})^{2}}} \nonumber \\ 
&\simeq& \left(\frac{1}{m_{\rm e}}+\frac{1}{\tilde{m}}\right) 
-\frac{1}{2}\left(\frac{1}{\tilde{m}}\right)^3\left(\frac{p}{\tilde{c}}\right)^2.
\end{eqnarray}
The approximation in the last line is valid when $p^2(\tilde{m} \tilde{c})^{-2} \ll 1$.
If we introduce new parameters $\tilde{\mu}=1/(1/m_{\rm e}+1/\tilde{m})$
and $\tilde{c}_{\mu}=(\tilde{m}/\tilde{\mu})^{3/2}\tilde{c}$, this becomes
\begin{equation}
\frac{1}{\mu(p)}\simeq \frac{1}{\tilde{\mu}}-\frac{1}{2}\left(\frac{1}{\tilde{\mu}}\right)^3 \left(\frac{p}{\tilde{c_{\mu}}}\right)^2,
\label{approxmass}
\end{equation}
which is equivalent to the expansion of the inverse of Eq. \eqref{mup} for $p^2(\tilde{\mu} \tilde{c}_{\mu})^{-2} \ll 1$. 
Therefore, $\tilde{\mu}$ and $\tilde{c_{\mu}}$ replace the mass $m$ and light speed $c$ in the Dirac equation, respectively.

The eigenvalues of the Dirac equation for a hydrogen atom are known as
\begin{equation}
E=\frac{mc^2}{
\sqrt{1+ \frac{\alpha^2}{\left( n-(j+1/2)+\sqrt{(j+1/2)^2-\alpha^2}\right)^2}}},
\label{DiracEq}
\end{equation}
where $j=\ell\pm1/2$ is the total angular momentum.
Because the reduced mass is expressed by Eq. \eqref{mup}, we substitute $\tilde{\mu}$,
$\tilde{c_{\mu}}$, and $\tilde{\alpha}=e^2/(4\pi\epsilon_{\rm b} \epsilon_0 \hbar \tilde{c_{\mu}})$ for $m$, $c$, 
and the fine-structure constant $\alpha$ in Eq. (\ref{DiracEq}). This leads to
\begin{equation}
\tilde{E}^{\rm Dirac}=\frac{\tilde{\mu}\tilde{c_{\mu}}^2}{
\sqrt{1+ \frac{\tilde{\alpha}^2}{\left( n-(j+1/2)+\sqrt{(j+1/2)^2-\tilde{\alpha}^2}\right)^2}}},
\label{DiracEq_NP}
\end{equation}
which takes account of the nonparabolicity under the condition $p^2(\tilde{m} \tilde{c})^{-2} \sim 0.1 \ll 1$.

The binding energies $E_{n,j}^{\rm Dirac}=\tilde{E}^{\rm Dirac}-\tilde{\mu}\tilde{c_{\mu}}^2$ of excitons with principal quantum numbers $n=1-30$ and angular momentum $\ell=0-5$ were calculated using Eq. \eqref{DiracEq_NP}, 
where $\ell=0,1,2,3,4,$ and 5 correspond to S-, P-, D-, F-, G-, and H-envelope states, respectively.
The values are summarized in Tables \ref{Diraccalc} and \ref{oddparity},
with notations using $n$(S,P,D,F,G,H)$_j$ for each state.
The tables also list the experimental values from this study and those from the literature \cite{Kazimierczuk2014,Stolz,highang,para,Krueger,Schone2016,Julian,para}.
The calculated binding energies were in good agreement with the measured values.

\subsection{Schr\"{o}dinger equation in momentum space}

\begin{figure}[t]
\centering
\includegraphics[width=8.5cm, clip]{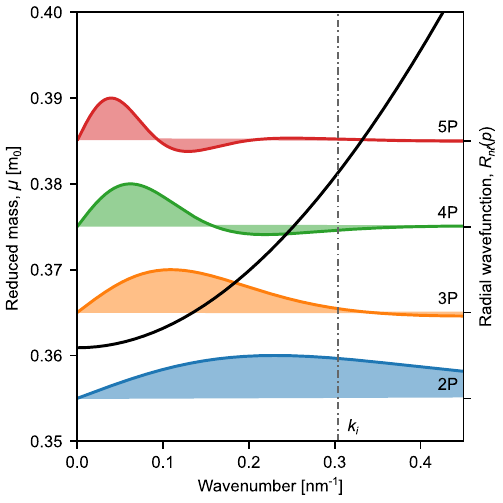}
\caption{
Radial wavefunctions of 2P$_{3/2}$, 3P$_{3/2}$, 4P$_{3/2}$, and 5P$_{3/2}$ states (offset vertically for clarity) calculated using Eq. \eqref{EqS10} as solutions of Schr\"{o}dinger equation for particles with a nonparabolic reduced mass.
Solid line: Wavenumber ($k$) dependence of the nonparabolic reduced mass $\mu(\hbar k)$ near the $k=0$ point, calculated with $\tilde{m^*}= 0.57027 m_0$, $\tilde{c^*}= 1.62755 \times 10^{5}$ m/s, and $T=0$ K. 
The dash-dotted line indicates the initial wavenumber, $k_{\rm i}$. 
}
\label{FigureS5}
\end{figure}

Following Ref. \cite{energycalc}, 
we consider a counterpart integral equation to the Schr\"{o}dinger equation, obtained by the Fourier transformation of its position representation,
that is,
\begin{equation}
\left(\frac{p^2}{2\mu(p)}-E \right)\Psi(E,p)=\frac{e^2}{4\pi\epsilon_{\rm b} \epsilon_0}\frac{\Gamma(1)}{2\pi^2\hbar}
\int d^3p' \frac{\Psi(E,p')}{|p'-p|^2}.
\end{equation} 
Finding negative energy eigenvalues $E(<0)$ for bound states is reduced to solving a Sturmian eigenvalue problem.
The solutions are expanded into series of products of Gegenbauer polynomials $C_{n_r}^{\ell+1}(x)$ with the radial quantum number
$n_r=n-\ell-1$, 
which leads to an integral equation for $radial$ functions called Coulomb Sturmian functions. 
Using the notation in Eq. (3.32) in Ref. \cite{energycalc}, the function entering the kernel of the integral equation is expressed as
\begin{eqnarray}
g_{n_r\ell}(p)=\Gamma \left(\ell+1\right) \sqrt{\frac{q_B n_r!}{\pi (n_r+2\ell+1)!}}  \nonumber\\
\times 
\frac{(4qp)^{\ell+1}}{(q^2+p^2)^{\ell+3/2}}C_{n_r}^{\ell+1}\left(\frac{q^2-p^2}{q^2+p^2}\right),
\end{eqnarray}
where $q_B=\hbar/a_B$ is defined using the Bohr radius $a_B= (4\pi\epsilon_{\rm b}\epsilon_0)\hbar^2/(e^2\tilde{\mu})$,
and $q=\sqrt{-2\mu(p)E}$ (which contains the nonparabolicity) is assumed for the renormalized energy.
Here, in analogy with the non-relativistic hydrogen case, we assume that $g_{n_r\ell}(p)$  forms an orthogonal set:
\begin{equation}
\delta_{n_r n_r'}= \int_0^\infty dp g_{n_r \ell}(p)g_{n_r' \ell}(p).
\label{orthogonal}
\end{equation} 
We searched for a value of $E$ with an accuracy of $10^{-8}$ eV to satisfy this relation for given sets of $n_r$ and $\ell$,
where the lower and upper bounds of integration were taken as $5\times 10^{-5}$ m$^{-1}$ and $1\times 10^{10}$ m$^{-1}$, respectively.

The radial part of the wavefunction is expressed as \cite{energycalc}
\begin{eqnarray}
R_{nl}(p)&=&2^{5/2}\Gamma\left(\ell+1\right) \sqrt{\frac{n_r! \, n}{\pi(n+\ell)!}}  \nonumber \\
&\times& \frac{q^{5/2} (4qp)^{\ell}}{(q^2+p^2)^{\ell+2}} C^{\ell+1}_{n_r} \left(\frac{q^2-p^2}{q^2+p^2}\right). 
\label{EqS10}
\end{eqnarray}
To approximate the effect of nonparabolicity, we substituted $q=\sqrt{-2\mu(p) E}$ using the numerically obtained $E$ values.
Figure \ref{FigureS5} shows the radial wavefunctions of the P$_{3/2}$ states with $n=2 - 5$, obtained using this approach. 
In the Figure, the nonparabolic band dispersion of the reduced mass $\mu(\hbar k)$ is shown by the black line.
The valid wavenumber range for the nonparabolic band is limited by the time origin ($t=t_0$) in Eq. \eqref{energy_time}, 
which corresponds to $k_{\rm i}=p_{\rm i}/\hbar=0.30 $ nm$^{-1}$. 
It is found that the wavefunctions with $n \ge 3$ reside mostly within $k_{\rm i}$. However,
as the nonparabolicity parameter is not necessarily the same outside this wavenumber range, 
we cannot guarantee the calculated binding energies for $n\leq2$ states. 
Nevertheless, good agreement is found between the measured and calculated 2P binding energies because the extrapolated band dispersion is within the anisotropic variation predicted by the DFT calculations [See Fig. \ref{FigureS4}(b)].

\begin{table*}[t]
\tabcolsep=6pt
\caption{
Comparison of quantum defect values ($\delta_j^{\rm Dirac}$ and $\delta_{n,\ell}$) from this study and $\delta_{n,\ell}$ in literature. 
$R$ represents the Rydberg energy. Numbers in parentheses are relative values based on $\delta_{n,\ell} \approx$ (128 meV)$(2 (\ell+1) R)^{-1}$ as derived in Eq. (24) of Ref. \cite{Julian}.
In Refs. \cite{Stolz,Stolz2022}, $E_n=R/(n-\delta_1-\delta_2/(n-\delta_1)^2-\cdots)^2$ was used instead of Eq. (2).
}
\label{quantumdefect}
\begin{tabular}{|c|c|c|c|c|c|c|c|}\hline
Dirac eq. & $R$ (meV) & $\delta_{1/2}^{\rm Dirac}$ & $\delta_{3/2}^{\rm Dirac}$ 
& $\delta_{3/2}^{\rm Dirac};\delta_{5/2}^{\rm Dirac}$
& $\delta_{5/2}^{\rm Dirac};\delta_{7/2}^{\rm Dirac}$ 
& $\delta_{7/2}^{\rm Dirac};\delta_{9/2}^{\rm Dirac}$
& $\delta_{9/2}^{\rm Dirac};\delta_{11/2}^{\rm Dirac}$\\ \hline \hline
This study & $87.3$ & $0.569$ & $0.215$ & $0.215;0.139$ & $0.139;0.103$ & $0.103;0.0821$ & $0.0821;0.0682$\\ \hline 
\noalign{\medskip} \hline
Eq. (2) & $R$ (meV) & $\delta_{\rm S}$ & $\delta_{\rm P}$ $[\delta_1]$ & $\delta_{\rm D}$ & $\delta_{\rm F}$ & $\delta_{\rm G}$ & $\delta_{\rm H}$ \\ \hline \hline
This study & $87.6\pm 0.1$ & $0.52$ & $0.17$ & $0.09$ & $0.06$ & 0.04 & 0.02\\ \hline 
a & $87.46198$ & & $[0.206]$ & & && \\\hline
b & $87.3\pm0.2$ & & $[0.225\pm0.02]$ & & &&\\\hline
c & 87 & $$ & $0.23$ & $$ & $$ &&\\ \hline
d  & 86.07 & $0.44$& $0.29$ & $0.16$ & $0.10$ &&\\ \hline
e & 86.04 & $0.57$ & $0.32$ & $0.19$ & $0.12$ &&\\ \hline
f & 86 & $0.65$ & $0.34$ &$0.18$ & $0.12$ && 0.075\\ \hline
g & 86 & $0.87$ & $0.43$ & $0.28$ & $0.21$ &&\\ \hline
h & 86 & $(4)$ & $(2)$ & $(1.3)$ & $(1)$ & (0.80) & (0.67)\\ \hline
i  & 86 & & 0.21 & & &&\\ \hline 
j & $82.40\pm0.05$ & $0.564\pm0.002$ & $0.310\pm0.003$ & $0.085\pm0.003$ & $0.185\pm0.007$ & & \\ \hline
k & $82.2\pm0.2$ & $0.468\pm0.005$ & $0.210\pm0.013$ & $-0.01\pm0.02$ & $0.12\pm0.06$ & & \\ \hline
l &   &  0.63 & 0.29 & 0.15 & 0.09 & &\\ \hline
\end{tabular}

a: Stolz 2021 \cite{Stolz}, b: Stolz 2022 \cite{Stolz2022}, c: Kazimierczuk 2014 \cite{Kazimierczuk2014},
d: Kr\"{u}ger 2021 \cite{Krueger}, e: Sch\"one 2016 \cite{Schone2016}, f: Heck\"otter 2017 (experiment) \cite{Scaling}, g: Heck\"otter 2017  (theory) \cite{Scaling},
h: Heck\"otter 2025 (relative value) \cite{Julian}, i: Versteegh 2021 \cite{PLE}, j: Rogers 2022 (incoherent model) \cite{Rogers}, k: Rogers 2022 (coherent model) \cite{Rogers}, l: Ertl 2026 (theory) \cite{Ertl2026}.
\end{table*}

\subsection{Comparison of quantum defects in present and previous works}

We calculated quantum defect values according to the Dirac approach: $\delta_j^{\rm Dirac}=(j+1/2)-\sqrt{(j+1/2)^2-\tilde{\alpha}^2}$.
The results are summarized in Table \ref{quantumdefect}. 
$\delta_j^{\rm Dirac}$ matches $\delta_{n,\ell}$ in Eq. (2) from the quantum mechanical framework, 
up to the order of $n^{-3}$ of the eigenenergies \cite{Jacobs}.
It is evident that our approach reproduces the trend of decreasing quantum defects with increasing $\ell$.
Furthermore, considering the simple description of the band dispersion using only two parameters in the present study, 
the semi-quantitative agreement with the values from previous studies \cite{Stolz,Stolz2022,Kazimierczuk2014,Krueger,Schone2016,Scaling,PLE,Julian,Rogers,Ertl2026} is remarkably good.
We note that the quantum defect value varies with the choice of Rydberg energy even when the same experimental data are used.
Indeed, by fitting the binding energies ($10\le n\le 25$) of the Dirac equation with Eq. (2),
we obtained $\delta_{\rm P} = 0.27$ and 0.33, when the Rydberg energy was fixed at $R=87$ and 86 meV, respectively.
Generally, the quantum defect values become smaller when a larger Rydberg energy is assumed, which explains our relatively small 
$\delta_{\ell}$ values for $R=87.6$ meV (in the fourth row of Table S4).

In previous theoretical studies \cite{Impact,Even,Ertl2022,Efield,Schone2016,Coupled,Scaling}, 
the effective-mass Hamiltonian
\begin{equation}
\mathcal{H}=\mathcal{H}_{\rm hydro}+\mathcal{H}_{d}^{\rm sph}+\mathcal{H}_d^{\rm cubic}+\mathcal{H}_{\rm ccc},
\end{equation}
was used to describe the excitons in Cu$_2$O.
The first term $\mathcal{H}_{\rm hydro}$ contains the kinetic energy and Coulomb interaction, similarly to the non-relativistic hydrogen case. 
The second and third terms lead to nonparabolicity. 
In the spherical approximation up to $\mathcal{H}_{d}^{\rm sph}$, $j$ remains a good quantum number, 
and labeling the states using $n$, $\ell$, and $j$ is appropriate. 
When the orbital angular momentum $\ell$ and hole spin are combined, degeneracy for different values of $j$ is lifted. 
The P-envelope states are split into $n$P$_{1/2}$ and $n$P$_{3/2}$ states. 
Although the $n$P$_{3/2}$ states are electric-dipole allowed,
the $n$P$_{1/2}$ states have $\Gamma_2^-\oplus\Gamma_5^-$ symmetries, for which electric-dipole transitions are forbidden.
The F-envelope states are split into $n$F$_{5/2}$ and $n$F$_{7/2}$ states, with both having electric-dipole allowed components.
The inclusion of the cubic term $\mathcal{H}_d^{\rm cubic}$ lowers the symmetry and makes $j$ not a good quantum number anymore. 
This even-parity operation mixes the states with the same parity.  
The fourth term, or the central-cell correction $\mathcal{H}_{\rm ccc}$, consists of the dielectric correction and electron-hole spin exchange interaction, 
both of which act primarily on the ground state with $n=1$. In particular, the exchange interaction acts only on the $\Gamma_5^+$ component of the even-parity exciton states, and the coupling strength decreases as $1/n^3$.

Our isotropic approximation in the Dirac approach corresponds to the inclusion of the spherical term and an angle-averaged cubic term.
As the exchange interaction is not considered, our predicted binding energies for the even-parity (S-, D-, and G-envelope) states can deviate from the observation, particularly for low $n$ states.
In contrast, good agreement is reached for the P- and F-envelope states observed in the one-photon transition spectra because the exchange interaction is zero for odd-parity states.
To conclude, our extremely simple description using the Dirac approach is suitable for calculating the binding energies of odd-parity giant Rydberg excitons.


\begin{thebibliography}{99}
\bibitem{note2}
The coefficient $\beta=(\Sigma_{i=1,2} \frac{3\alpha_i}{8E_{{\rm LO}i}})/(1+\Sigma_{i=1,2}\frac{\alpha_i}{6})$ was calculated based on the generalized 
Feynman polaron model \cite{polaron,Ohyama}.
In calculating the coupling constants $\alpha_i=e^2 \sqrt{\tilde{m^*}/(1+\Sigma_{i=1,2}\frac{\alpha_i}{6})/(2 E_{{\rm LO}i})} 
(1/\epsilon_{{\rm b}i}-1/\epsilon_{{\rm s}i})/(4\pi\epsilon_0 \hbar)$, we approximated $1+\Sigma_{i=1,2}\frac{\alpha_i}{6}$ by 1 
because $\Sigma_{i=1,2}\frac{\alpha_i}{6}\ll1$.
We used dielectric parameters ($\epsilon_{{\rm s}1}=7.50$ \cite{dielectric}, $\epsilon_{{\rm b}1}=\epsilon_{{\rm s}2}=7.27$, $\epsilon_{{\rm b}2}=6.65$) 
extracted from the Lyddane-Sachs-Teller relation and longitudinal optical phonon energies $E_{\rm LO1}=19.1$ meV and $E_{\rm LO2}=82.1$ meV \cite{Eph}.
\bibitem{note3}
The data points below 5 K were excluded in the fit because microwave heating occurs \cite{Konishi}.
\end{thebibliography}
\end{document}